\documentclass[conference,a4paper]{IEEEtran}
\IEEEoverridecommandlockouts

\usepackage{cite}
\usepackage{amsmath,amssymb,amsfonts}
\usepackage{algorithmic}
\usepackage{graphicx}
\usepackage{textcomp}
\usepackage{xcolor}
\usepackage{footnote}
\usepackage{hyperref}
\usepackage{xcolor}
\usepackage{tabu}
\usepackage{makecell}
\usepackage{float}
\usepackage{adjustbox}
\usepackage{multirow}
\usepackage{booktabs}
\usepackage[inline]{enumitem}
\usepackage{lipsum}
\usepackage[]{caption}
\usepackage{subcaption}
\usepackage{soul}
\usepackage{flushend}

\usepackage[a4paper,top=1.90cm,bottom=2.9cm,left=1.57cm,right=1.57cm]{geometry} 

\definecolor{boristext}{rgb}{0.22, 0.44, 0.88}
\definecolor{boriscomments}{rgb}{0.88, 0.04, 0.04}
\definecolor{boristochange}{rgb}{0.2, 0.8, 0.8}

\begin{document}

\title{IEEE 802.11be Multi-Link Operation: When the Best Could Be to Use Only a Single Interface}

\author{\IEEEauthorblockN{ \'Alvaro L\'opez-Ravent\'os}
\IEEEauthorblockA{\textit{Dept. Information and Communication Technologies} \\
\textit{Universitat Pompeu Fabra (UPF)}\\
Barcelona, Spain \\
alvaro.lopez@upf.edu}
\and
\IEEEauthorblockN{ Boris Bellalta}
\IEEEauthorblockA{\textit{Dept. Information and Communication Technologies} \\
\textit{Universitat Pompeu Fabra (UPF)}\\
Barcelona, Spain \\
boris.bellalta@upf.edu}}

\maketitle


\begin{abstract}
The multi-link operation (MLO) is a new feature proposed to be part of the IEEE 802.11be Extremely High Throughput (EHT) amendment. Through MLO, access points and stations will be provided with the capabilities to transmit and receive data from the same traffic flow over multiple radio interfaces. However, the question on how traffic flows should be distributed over the different interfaces to maximize the WLAN performance is still unresolved. To that end, we evaluate in this article different traffic allocation policies, under a wide variety of scenarios and traffic loads, in order to shed some light on that question. The obtained results confirm that congestion-aware policies outperform static ones. However, and more importantly, the results also reveal that traffic flows become highly vulnerable to the activity of neighboring networks when they are distributed across multiple links. As a result, the best performance is obtained when a new arriving flow is simply assigned entirely to the emptiest interface.
\end{abstract}



\section{Introduction}


Since its appearance back in the late '90s, Wi-Fi has been continuously innovating to adapt its performance to the new user demands. Nowadays, the appearance of modern applications are pushing wireless local area networks (WLANs) to their performance limits again. For instance, remote office, cloud gaming and virtual and augmented reality (VR\&AR) are straight-forward examples that not only demand high-throughput but reliable and low-latency communications \cite{carrascosa2020cloud}. To that end, in may 2019, the Institute of Electric and Electronic Engineers (IEEE) established a Task Group (TG) to address and design a new physical (PHY) and medium access control (MAC) amendment, known as IEEE 802.11be Extremely High Throughput (EHT).

The IEEE 802.11be EHT seeks to further increase the throughput performance, reduce the end-to-end latency and increase the reliability of communications  \cite{adame2019time}. To do so, different technical features have been suggested in both PHY and MAC layers~\cite{lopez2019ieee}. Regarding PHY layer, TGbe propose the complementary adoption of the 6 GHz band, empowering the use of wider bandwidth channels up to 320 MHz, additionally to a new 4096 QAM high-order modulation (i.e., 12 bits per symbol), as immediate approaches to increase Wi-Fi peak-throughput. Besides, the multiple-input multiple-output (MIMO) capabilities are also being revised to upgrade them to support up to 16 spatial streams, introducing also an implicit channel sounding procedure. Despite those enhancements, it is not in the PHY, but in the MAC layer where we can find the most disruptive updates. We refer to the adoption of multi-link communications, which represents a paradigm shift towards concurrent transmissions. Although under the multi-link label we find the multi-AP coordination and the multi-band/multi-channel operation features, this article is focused on the analysis of the latter one.

Upon its current version, the IEEE 802.11 standard already defines two MAC architectures for supporting the multi-band/multi-channel operation. However, both designs present a common limitation: MAC service data units (MSDUs) belonging to the same traffic flow can not be transmitted across different bands~\cite{FangEHT2019}. As a result, stations are tied to a single band, preventing a dynamic and seamless inter-band operation. That is, one link is selected between transmitter and receiver to carry out data exchanges, remaining the other unused~\cite{fadlullah2017multi, kyasanur2005routing, singh2011green, choi2003multi}. To benefit from the fact that modern APs and stations incorporate dual, or even, tri-band capabilities, TGbe is working on developing the multi-link operation\footnote{Throughout this paper, we will refer to the multi-band/multi-channel operation feature as the MLO, following the notation of the TGbe.} (MLO) framework to allow concurrent data transmission and reception in different frequency channels/bands.

MLO allows APs and stations to exploit the fact of having multiple radio interfaces to transmit and receive data. However, how to use them to maximize the WLAN performance, i.e., how to properly distribute the traffic across the multiple interfaces, is still an open question. Therefore, in this paper, we tackle such question by considering different traffic balancing policies. The obtained results show the advantages of coordinating the available interfaces through the MLO framework, as well as provide some insights on how to distribute the traffic across them. It is worthy to anticipate our conclusion that distributing the traffic over multiple interfaces may not be always the best solution.



\section{Multi-link operation: an overview}\label{MLO_overview}

The introduction of MLO is a breaking point for Wi-Fi, as its adoption represents a paradigm shift towards multi-link communications. However, this implementation involves new challenges in terms of APs' and stations' architecture design, transmission modes, channel access, and management. 

To enable a concurrent operation across multiple interfaces, the existing multi-band MAC architecture have been exhaustively revised. In this context, TGbe introduces the concept of multi-link capable device (MLD), which consists of a single device with multiple wireless PHY interfaces that provides a unique MAC instance to the upper layers. Such implementation is achieved by dividing the MAC sub-layer in two parts~\cite{FangEHT2019_2}. First, we find the unified upper MAC (U-MAC), which is the common part of the MAC sub-layer for all the interfaces. Apart from performing link agnostic operations such as A-MSDU aggregation/de-aggregation and sequence number assignment, the U-MAC implements a queue that buffers the traffic received from the upper-layers. That is, traffic awaits in the U-MAC before it is assigned to a specific interface to be transmitted. Also, the U-MAC provides some management functions such as multi-link setup, association and authentication. On the other hand, there is the independent low MAC (L-MAC), an individual part of the MAC sub-layer for each interface that performs link specific functionalities. There, we find the link specific EDCA queues (one for each access category) that hold the traffic until its transmission. Also, procedures such as MAC header and cyclic redundancy check (CRC) creation/validation, in addition to management (e.g., beacons) and control (e.g., RTS/CTS and ACKs) frame generation are implemented at the L-MAC~\cite{LevyEHT2020}. 

This two-tier MAC implementation enables frames to be simultaneously transmitted over multiple links, as the U-MAC performs the allocation and consolidation of MPDUs when acting as transmitter and receiver, respectively. Also, it enables seamless transitions between links minimizing the access latency, and addressing an efficient load balancing.

At a link level (i.e., in the L-MAC), channel access takes place. In current proposals, TGbe defines different channel access methods accordingly to two different transmission modes: asynchronous and synchronous. First, there is the asynchronous transmission mode. Under this operational mode, a MLD can transmit frames asynchronously on multiple links, while keeping for each one its own channel access parameters (e.g., contention window (CW), arbitration inter-frame spacing number (AIFSN), etc.). That is, each link has its own primary channel, independent of the others. Also, this transmission mode allows the simultaneous transmission and reception (STR) capability over multiple links, enabling concurrent uplink and downlink communications. Such implementation has been proved to minimize latency \cite{naikcan}, while maximizing the throughput \cite{naribole2020simultaneous}. Ideally, the asynchronous mode should be selected as operational scheme, however, a MLD may not be STR-capable due to the in-device coexistence (IDC) interference. This issue is caused by an excessive power leakage between interfaces that do not have sufficient frequency separation in their operating channel (e.g., two channels in the 5~GHz band). As a result, IDC prevents frame reception on one interface, during an ongoing transmission on the other interface. To avoid the IDC issue, TGbe proposes the synchronous mode, which relies on synchronized frame transmissions across the available interfaces. With that, APs or stations are prevented to perform an STR operation, avoiding IDC problems but at the cost of a lower network throughput~\cite{MurtiEHT2019,PatilEHT2019}. Under this operation mode, the channel access can be performed either following the single primary channel (SPC) or the multiple primary channel (MPC) methodology. Basically, the SPC performs contention on a unique channel, whereas in the MPC contention is performed on all channels. Either using the SPC or the MPC method, if one channel wins contention the others are checked during a PCF inter-frame space (PIFS) time to see if they can be aggregated. As a result, channels in idle state are aggregated and a synchronous transmission starts~\cite{JangEHT2019}. It is worth mention that AP MLDS may change its transmission modes (e.g., asynchronous to synchronous, and vice versa) at any time. Devices operating in a synchronous mode are referred to as constrained MLDs, or non-STR MLDs, since they are not allowed to transmit through an idle interface at the same time they are receiving through another.

Regarding MLO feature management, the TGbe proposes the multi-link setup process. Instead of designing new management frames, TGbe agreed to reuse the current association request/response frames by adding an extra multi-link element or field. That is, the setup (i.e., MLO capability exchange) is performed jointly with the association mechanism. Then, through the multi-link element, AP MLDs and non-AP MLD negotiate and establish their subsequent operation scheme by exchanging their capabilities. For instance, they exchange information regarding the number of supported links, their ability to perform STR, their transmission operation (asynchronous or synchronous), and other per-link information (e.g., frequency band, supported bandwidth, number of spatial streams, etc.)~\cite{JangEHT2019_2, JangEHT2019_3}. To reduce overhead, the described multi-link setup process is proposed to be performed only on a single link, which, indeed, will be the lowest in frequency due to propagation constraints. 

The interface usage negotiation is performed by measuring link qualities at all interfaces. That is, those receiving a quality value above the clear channel assessment (CCA) threshold are set as enabled, while disabled otherwise. Hence, for each enabled interface, we will have an enabled link. It is worth mention that, as users may keep themselves mobile, any link listed as disabled may be added afterwards to the set of enabled links, by requesting a re-setup.


\section{System model}\label{Sys_model}

\subsection{Node placement}\label{subsec:node_placement}

We consider a set of IEEE 802.11be MLO-capable WLANs with $N$~APs and $M$~stations. Over a given area, APs are placed uniformly at random. However, we only accept the generated scenario as valid if the inter-AP distance is higher than 5~m to avoid unrealistic overlaps. Otherwise, we discard it and generate a new one. Then, we decide the number of stations that will be associated to each AP, placing them around it at a distance $d$ within the interval [$1$-$8$]~m and an angle $\theta$ between [0-2$\pi$], both selected uniformly at random.

Regardless of its type, APs and stations are configured with 3 wireless interfaces (i.e., $i_1$, $i_2$ and $i_3$), each one using 2 SU-MIMO spatial streams. Additionally, we consider that each interface operates at a different frequency band. That is, for each AP MLD interface, one channel $c$ is selected uniformly at random from the set of channels of each band, where $\mathcal{C}_b$ is the set of available channels of band $b$. Thus, ${c_{i_1}\hspace{0.5pt}\in\hspace{0.5pt}\mathcal{C}_{2.4}}$, ${c_{i_2}\hspace{0.5pt}\in\hspace{0.5pt}\mathcal{C}_{5}}$ and
${c_{i_3}\hspace{0.5pt}\in\hspace{0.5pt}\mathcal{C}_{6}}$. The different set of channels are detailed in Table~\ref{table:sim_params}. Note that depending on the band, different channel widths are considered.

\subsection{MLO capabilities}
The stations' set-up, as well as the establishment of the enabled interfaces, is performed through the 2.4~GHz link. Link qualities are exchanged, and therefore, interfaces that meet the quality criteria (i.e., above CCA threshold) are set as enabled, while disabled otherwise. In essence, this setup relies on a simplified version of the MLO setup process described in Section~\ref{MLO_overview}. Additionally, we assume that all nodes perform an asynchronous channel access for each of their enabled interfaces, while their default policy is set as \textit{Multi Link Same Load to All interfaces} (MLSA) (see Section~\ref{subsec:policies}), which will be our baseline evaluation policy unless stated differently. 

\begin{figure*}[t]
    \centering
    \includegraphics[width=0.95\linewidth]{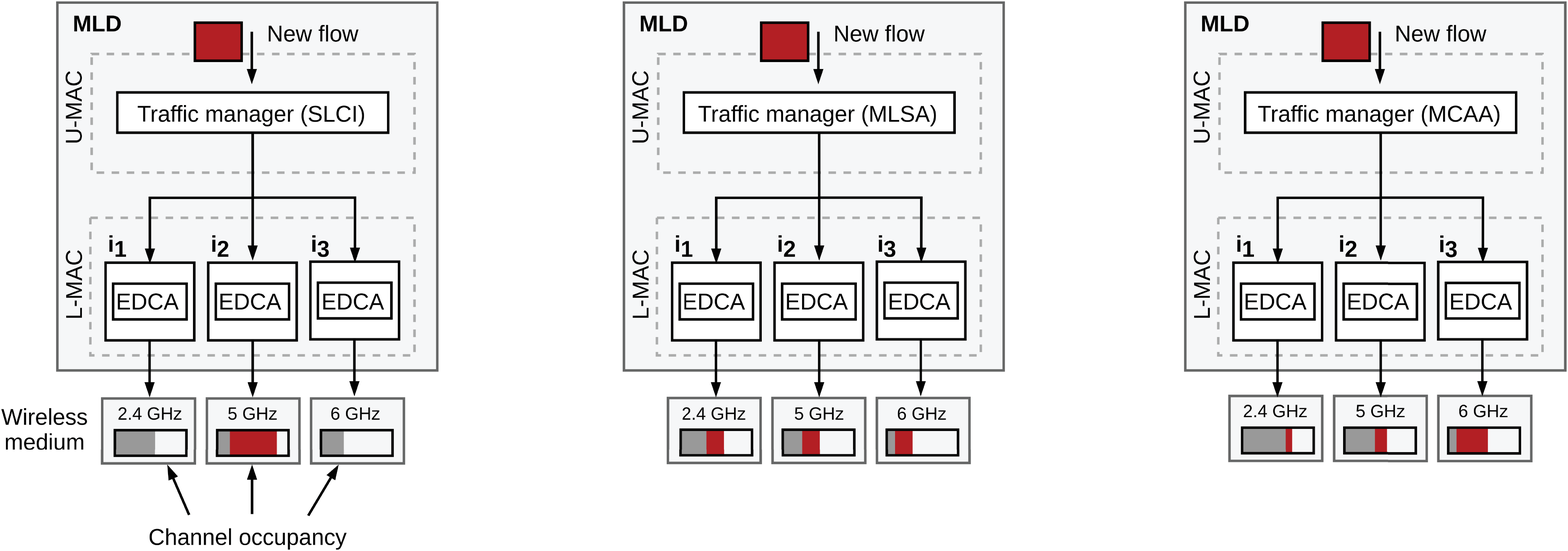}
    \caption{Policy implementation. From left to right: SLCI, MLSA and MCAA representation. Under the wireless medium representation, the gray area corresponds to the channel occupancy already seen by the MLD at the given interface, which can be from its neighbor APs, as well as its own ongoing flows.}
    \label{fig:FigOverview}
\end{figure*}

\subsection{Channel model and data rate selection}

Path loss is characterized following the IEEE 802.11ax enterprise model for a single floor environment. The path loss between an station $i$ and AP $j$ is given by
\begin{equation*}\label{eqn:pathloss}
\centering
\small
\begin{split}
    \text{PL} (d_{i,j}) &= 40.05 + 20\log_{10}\left(\frac{f_c}{2.4}\right) + 20\log_{10}(\text{min}(d_{i,j},d_{bp}))\\
    &+(d_{i,j}>d_{bp})\cdot35\log_{10}\left(\frac{d_{i,j}}{d_{bp}}\right) + 7W_{i,j}
\end{split}
\end{equation*}
where $f_c$ is the AP's carrier frequency in GHz, $d_{i,j}$ is the distance between station $i$ and AP $j$ in meters, $d_{bp}$ is the breaking point distance in meters, and $W_{i,j}$ are the number of traversed walls. We set the breaking point distance to 5~m and the number of traversed walls to~4. Note that the resulting propagation losses are given in dB.

The Modulation and Coding Scheme (MCS) used by each interface is selected according to the signal-to-noise ratio (SNR) between the AP and the stations. For instance, an station's interface~$i$, with 2 spatial streams and a SNR of 11~dB, will achieve a data rate of 243.8~Mbps, using modulation 1024-QAM with a coding rate of 5/6 and guard interval of 3.2~$\mu$s in a 20~MHz channel.

\begin{table}[t]
   \small 
   \centering 
   \caption{Evaluation setup}
   \resizebox{\linewidth}{!}{
       \begin{tabular}{ll} 
       \toprule
       \textbf{Parameter} & \textbf{Description}\\ 
       \midrule
       Carrier frequency & Depends on channel selection\\
       $\mathcal{C}_{2.4}$ channel set & 1 (20 MHz), 6 (20 MHz), 11 (20 MHz)\\
       $\mathcal{C}_{5}$ channel set & 38 (40 MHz), 46 (40 MHz), 58 (80 MHz)\\
       $\mathcal{C}_{6}$ channel set & 55 (80 MHz), 71 (80 MHz), 15 (160 MHz)\\
       System bandwidth & 60 MHz/160 MHz/320 MHz\\
       AP/STA TX power & 20/15 dBm\\
       Antenna TX/RX gain & 0 dB\\
       CCA threshold & -82 dBm\\
       AP/STA noise figure & 7 dB\\
       \makecell[l]{Single user \\ spatial streams} & 2\\
       MPDU payload size & 1500 bytes\\
       Path loss & Same as \cite{lopez2020concurrent}\\
       Avg. flow duration & T$_{\text{on}}=$ 1 s\\
       \makecell[l]{Avg. flow \\ interarrival time}  & T$_{\text{off}}=$ 3 s\\
       \makecell[l]{Min. contention \\ window} & 15\\
       Packet error rate & 10\%\\
       Simulation time & 120 s  (1 per simulation)\\
       Number of simulations & 100 (per evaluation point) \\
       \bottomrule
       \end{tabular}}
       \label{table:sim_params}
\end{table}

\subsection{Traffic generation and CSMA operation}\label{subsec:traffic}

Only downlink traffic is considered. The traffic directed to each station is modeled as an on/off Markovian model. In the on period, the AP receives a Constant Bit Ratio (CBR) traffic flow of $\ell$~Mbps, whereas zero during the off period. Both on and off periods are exponentially distributed with mean duration T$_{\text{on}}$ and T$_{\text{off}}$, respectively. Either T$_{\text{on}}$ and T$_{\text{off}}$ have been set to resemble a mixture of different types of traffic as in real scenarios. The value of the two parameters is presented in Table~\ref{table:sim_params}. We refer to the traffic generated during the on period as a traffic flow, and to $\ell$ as the required bandwidth.

Regarding the CSMA/CA operation, it follows the abstraction presented in~\cite{lopez2020concurrent}, which considers the aggregate channel load observed by each AP to calculate the airtime that can be allocated to each flow. Although the considered abstraction does not capture low-level details of the PHY and MAC layers operation, it maintains the essence of the CSMA/CA: the 'fair' share of the spectrum resources among contending APs and stations. In addition, this presented abstraction allows us to simulate larger networks, and larger periods of time, and so study the network performance at the flow-level.

\subsection{Traffic allocation policies}\label{subsec:policies}

When a new traffic flow arrives at the AP, the traffic manager determines how it is distributed over the different enabled interfaces of each station. The three traffic allocation policies considered in this work are:

\begin{itemize}
    \item \textbf{Single Link Less Congested Interface (SLCI):} upon a new flow arrival, pick the less congested interface and assign it to the incoming flow.
    \item \textbf{Multi Link Same Load to All interfaces (MLSA):} upon a new flow arrival, distribute the incoming traffic flow equally between all the enabled interfaces of the receiving station. That is, let $\ell$ be the required bandwidth of the incoming flow, and $N_{\rm int}$ the number of enabled interfaces in the destination station. This is, traffic allocation per interface is given by: $\ell_i = \ell / N_{int}$, with~$\ell_i$ the bandwidth allocated to interface~$i$.
    \item \textbf{Multi Link Congestion-aware Load balancing at flow arrivals (MCAA):} upon a new flow arrival, distribute the incoming traffic flow accordingly to the channel occupancy observed at the AP considering the enabled interfaces at the receiving station. That is, let $\rho_i$ the percentage of available (free) channel airtime at interface~$i$. Then, the required bandwidth allocated to interface $i$ is given by $\ell_{i \in \mathcal{J}}=\ell \frac{\rho_i}{\sum_{ \forall j \in \mathcal{J}}{\rho_j}}$, with $\mathcal{J}$ the set of enabled interfaces at the target station.
    %
\end{itemize}


While MLSA allocates traffic regardless of the congestion level of each interface, the other policies take it into account. When a new flow arrives, SLCI evaluates the current congestion level of each interface to identify the emptiest one, and so, it assigns the whole traffic flow to it. Then, we find MCAA, which tries to maximize the spectrum used by balancing the traffic allocated to the different interfaces taking into account their current occupancy. Under the MCAA, for instance, a flow with a bandwidth requirement of 10 Mbps, and capable to be distributed across three interfaces with $\rho_1=0.2$, $\rho_2=0.6$, and $\rho_3=0.5$, will be split as follows: 1.54~Mbps allocated to the interface~1, 4.61~Mbps to the interface~2, and 3.85~Mbps to interface~3. Although this approach seems very convenient, traffic may become more vulnerable to contention with external networks, due to the fact of being spread across multiple interfaces. Figure~\ref{fig:FigOverview} shows the different policies presented. 


\begin{figure*}[t]
    \centering
    \subcaptionbox{\small \label{fig:toy_topo}}[0.325\linewidth]{\includegraphics[width=\linewidth]{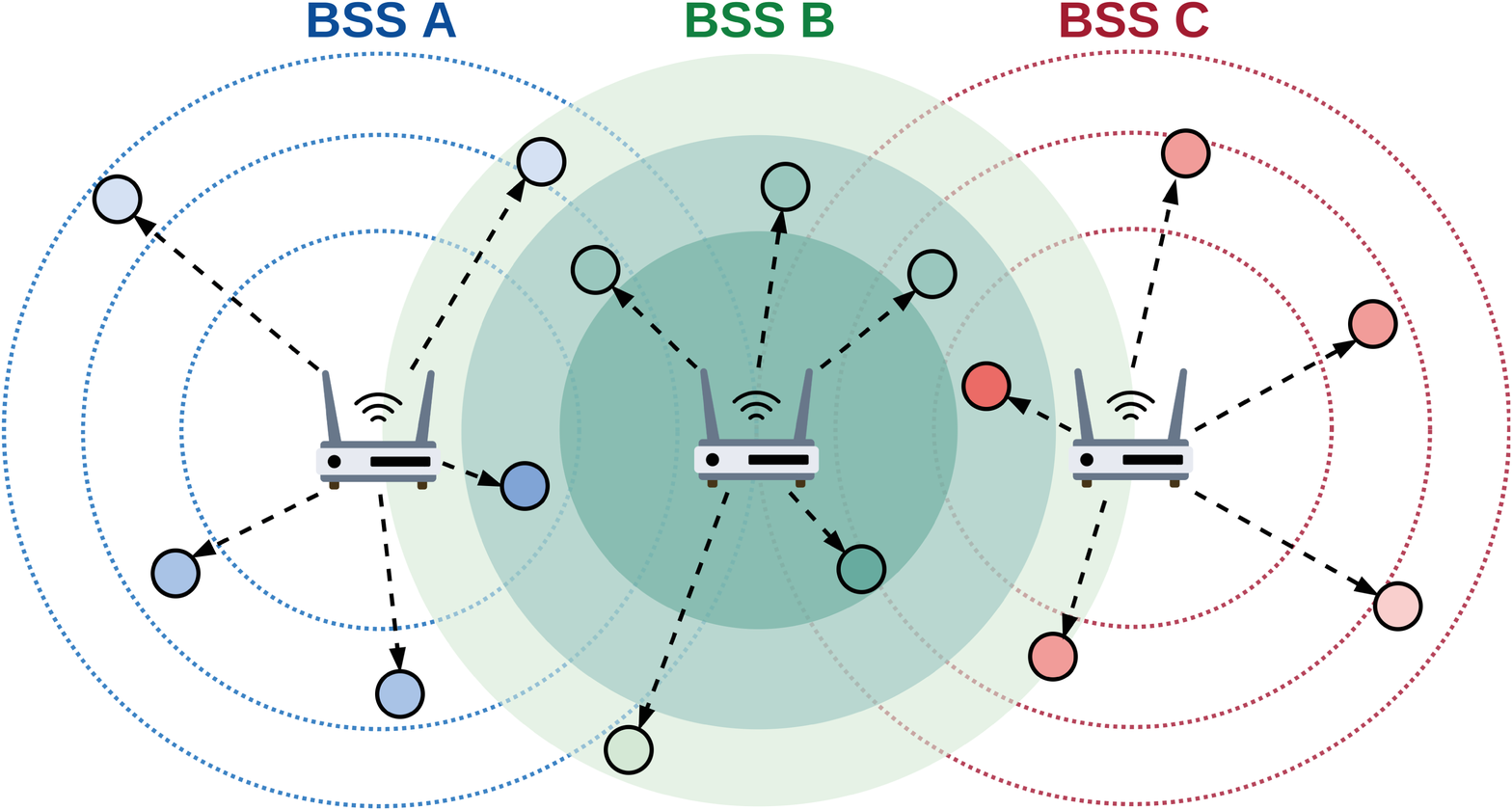}}\hfill%
    \subcaptionbox{\small \label{fig:toy_occ}}[0.325\linewidth]{\includegraphics[width=\linewidth]{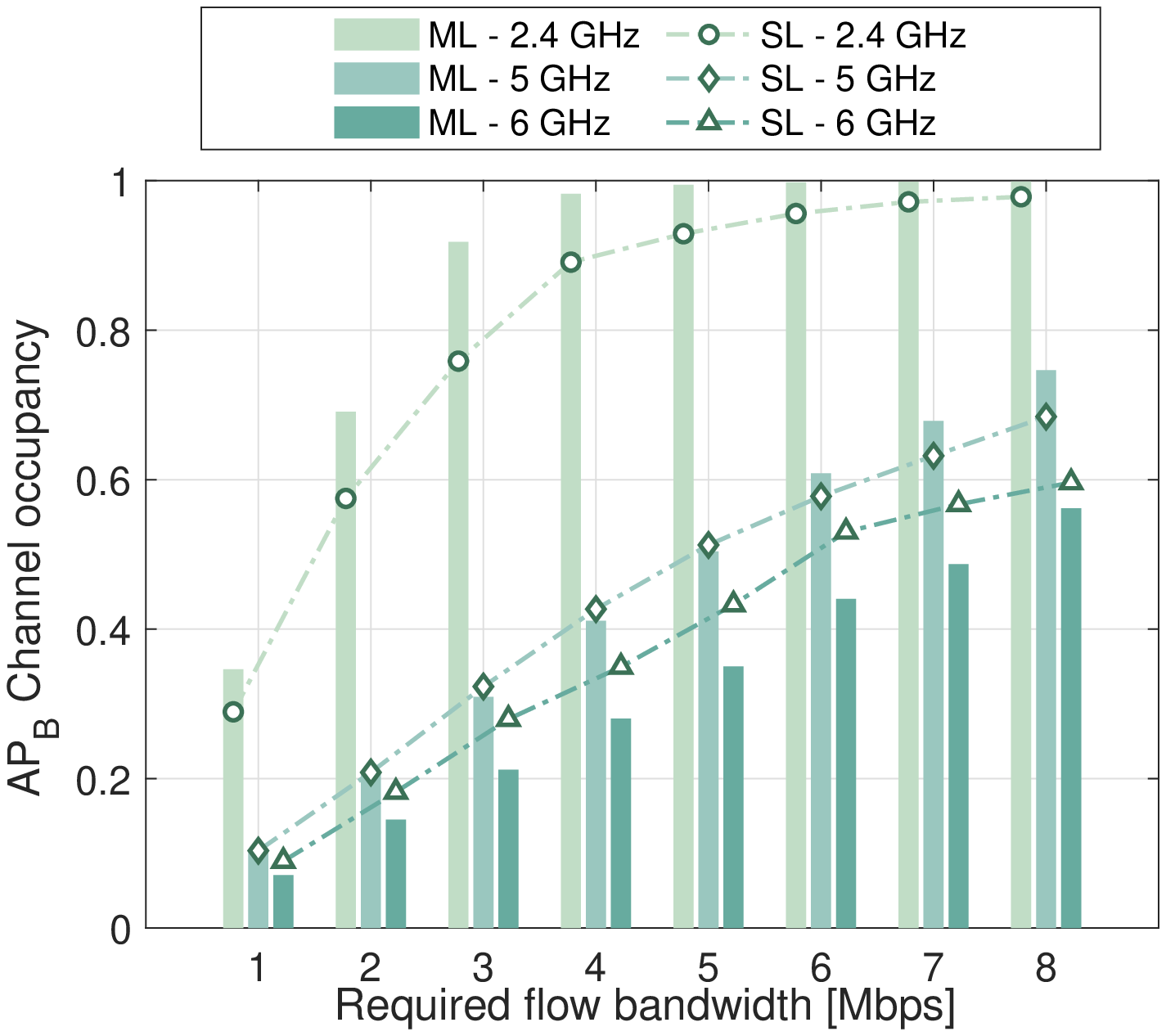}}\hfill%
    \subcaptionbox{\small \label{fig:toy_throughput}}[0.325\linewidth]{\includegraphics[width=\linewidth]{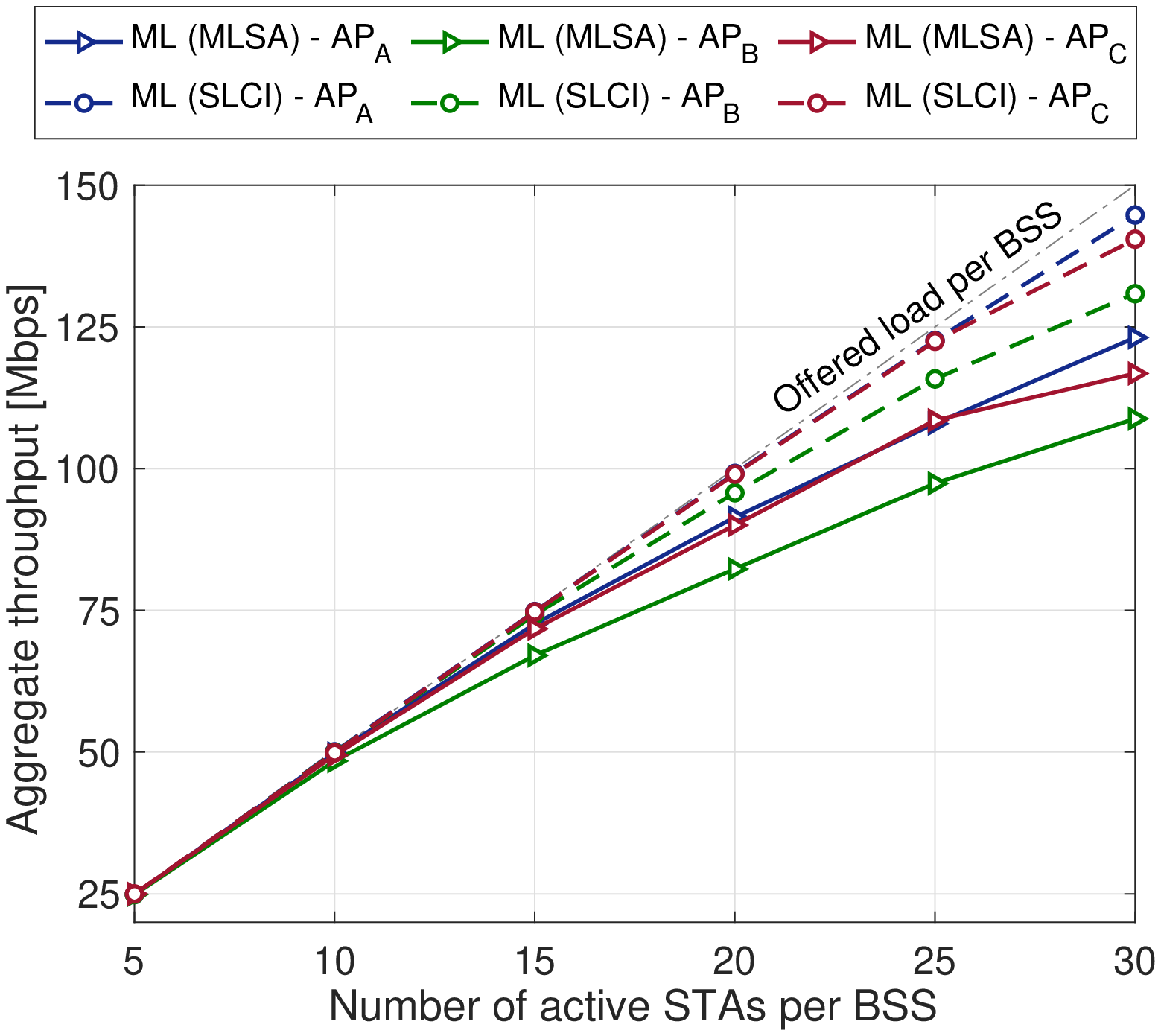}}
    \caption{Controlled scenario evaluation. From left to right: a) scenario representation, b) average channel occupancy for AP B, and c) aggregate throughput performance. The high, medium and low shaded areas represent the operation range for the 6~GHz, 5~GHz and 2.4~GHz bands, respectively.}\label{fig:1}
\end{figure*}

\section{Performance evaluation}\label{P_evaluation}

To assess the performance evaluation of the MLO, we use the Neko\footnote{The Neko simulation platform can be found in GitHub at: \url{https://github.com/wn-upf/Neko}} simulation platform.

\subsection{A controlled scenario}

First, we focus our evaluation on the comparison between a MLO-capable deployment against a traditional multi-band single link (SL). To do so, we consider a controlled scenario to pinpoint the potential benefits about the introduction of MLO, while identifying some pitfalls without the complex interactions that random deployments may originate. Figure~\ref{fig:toy_topo} depicts a typical controlled WLAN deployment, in which three basic service sets (BSSs) are placed inline. Under such scenario, we assess the network behavior under two different cases: \begin{enumerate*}[label=\roman*)] \item high traffic demands, and \item high station density \end{enumerate*} deployments. For each evaluation point considered in each case, we simulate 100 scenarios in which stations' positions are newly generated following the stated in Section~\ref{subsec:node_placement}.

To begin with, we characterize the network performance for different traffic loads. In this regard, we fixed the average number of associated stations per AP to 20, while increasing the required bandwidth for each flow from 1~Mbps to 8~Mbps. That is, 800 total simulated scenarios, as we evaluate 8 traffic load ranges for 100 different scenarios. Besides, the whole deployment is considered to be either multi-band SL (i.e.,~3~SL~APs), or MLO-capable (i.e.,~3~AP~MLDs). In both scenarios, the channel configuration is the same for the 3~APs, whereas the other simulation parameters are kept as presented in Table~\ref{table:sim_params}. It is worth mention that when using the multi-band SL approach, stations select the interface to be attached to in a random fashion from the set of enabled interfaces. For MLO, the configuration of each interface is set as stated in Section~\ref{subsec:node_placement}, and both APs and stations use baseline MLSA policy.

Figure~\ref{fig:toy_occ} shows the experienced channel occupancy by AP$_\text{B}$. We observe that AP$_\text{B}$ suffers from the well-known flow-in-the-middle effect in its 2.4~GHz interface, due to AP$_\text{A}$ and AP$_\text{C}$ keeping the wireless medium in busy state for most of their time. Consequently, traffic bandwidth requirements in this interface are only met for low level demands in both approaches, since AP$_\text{B}$ will start to drop data traffic as soon as traffic load increases. On the contrary, such effect is alleviated in 5~GHz and 6~GHz links as a consequence of nodes' spatial distribution, which provides all APs with two downlink contention-free links (i.e., only the 2.4GHz link satisfies the energy detection threshold). Although such condition may benefit the MLO capability implementation, in this case it is wasted by not using a  congestion-aware traffic allocation policy to leverage both contention-free links. Such issue has a negative impact over AP$_\text{B}$ performance, as its experienced occupation is nearly the same either when using both multi-band SL or MLO approaches. That is, there is no difference between spreading in the same proportion the traffic over the different links through MLO, than balancing and/or steering clients across links, like current multi-band SL deployments do. Thus, it is easy to conclude that without a proper traffic allocation policy, MLO underperforms. 

Similarly, Figure~\ref{fig:toy_throughput} shows the aggregate throughput of each AP when the number of stations per BSS increases. Differently from the previous case, we set now that the required bandwidth for each generated flow~${\ell}$ is selected uniformly from the 2~Mbps to 8~Mbps interval (i.e., ${\ell}\sim\textit{U}[2,8]$). The other simulation parameters are kept as previously. Now, we introduce the SLCI congestion-aware policy presented in Section~\ref{subsec:policies} to provide a comparison between a non-congestion and a congestion aware policy. From the figure, we observe how the MLSA policy struggles to cope even with 15 stations per BSS. For instance, AP$_\text{B}$ already registers throughput losses around 12\% for 15 stations. On the contrary, the SLCI assisted MLO operation supports a higher number of stations, while AP$_\text{B}$ only experiences a 5\% throughput losses when considering 20 stations. Additionally, we observe that the flow-in-the-middle effect is reduced by the implementation of a congestion-aware policy. For instance, we observe that AP$_\text{B}$'s aggregate throughput is reduced 10\% in comparison to its neighbors when considering MLSA for 20 stations, whereas only a 4\% when considering SLCI.

\subsection{Random deployments}

In order to get more insights with respect to the performance of the different policies presented in Section~\ref{subsec:policies}, we conduct a set of simulations in random generated scenarios. Similarly to the previous case, we address two different cases: \begin{enumerate*}[label=\roman*)] \item high user demands, and \item high AP density \end{enumerate*} deployments. Both cases are conducted over an area of 45x45$\text{m}^2$, with nodes being placed and configured as stated in Section~\ref{Sys_model}. Besides, we consider that both APs and stations within the evaluation area implement the same policy. Other simulation conditions are kept as stated in Table~\ref{table:sim_params}. For each evaluation point considered in each case, we simulate 100 different scenarios in which APs and stations' positions are generated as stated in Section~\ref{subsec:node_placement}.

\begin{figure}[t]
    \centering
    \includegraphics[width=0.7\columnwidth]{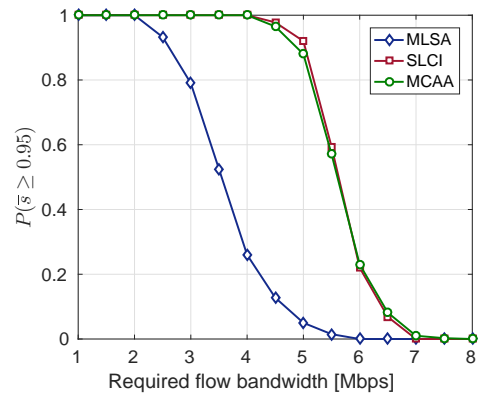}
    \caption{Prob. of average satisfaction}\label{fig:load_prob}
\end{figure}

First, we conduct the evaluation with respect to stations traffic requirements. For this purpose, we locate 10~AP~MLDs within the considered area, as well as a random number of stations (i.e., $M\sim U[15,25]$) for each AP~MLD. Then, we carry out simulations by gradually increasing the required bandwidth~${\ell}$ of each incoming flow from 1~Mbps to 8~Mbps. To evaluate the performance of the network we use the satisfaction $s$ metric. We define the satisfaction as the ratio between the required airtime to achieve the required flow bandwidth, and the actual amount that can be allocated. This metric is calculated for each AP, and then, ~$\overline{s}$ indicates the average satisfaction of the network computed as the sum of the average satisfaction experienced by each AP divided by the total number of APs in the WLAN. Figure~\ref{fig:load_prob} shows the probability that the whole WLAN achieves an average satisfaction~$\overline{s}$ greater or equal than 95\%. As expected, the baseline MLSA operation is outperformed by the rest of policies, due to the fact that traffic allocation is no longer allocated blindly, but following a congestion-aware approach. In this regard, we find that both SLCI and MCAA congestion-aware policies are capable to deal with higher traffic loads without losing performance. For instance, for~${\ell~=~5~\text{Mbps}}$, SLCI and MCAA are able to provide excellent satisfaction values in 92\% and 88\% of the scenarios evaluated, respectively, whereas MLSA only manages to get good results in 5\% of them. Finally, notice that SLCI works better for medium load ranges, while MCAA works better for high loads as its ability to distribute traffic across the different interfaces results in slightly higher satisfaction values. 

\begin{table}[t!!!]
    \centering 
    \caption{Traffic allocation efficiency}
    \resizebox{0.75\linewidth}{!}{
    \begin{tabular}{c c c c}
        \toprule
        \multirow{2}{*}{\makecell[c]{bandwidth \\ req. per flow}} & \multicolumn{3}{c}{Policy}\\
        \cmidrule{2-4}
        & MLSA & SLCI & MCAA \\
        \midrule
        2~Mbps & 0.996 & 1.00 & 1.00 \\
        4~Mbps & 0.925 & 0.989 & 0.985 \\
        6~Mbps & 0.830 & 0.931 & 0.930 \\
        8~Mbps & 0.750 & 0.833 & 0.842 \\
        \bottomrule
    \end{tabular}}
    \label{table:bw_eff}
\end{table}

To further examine the obtained results, we show in Table~\ref{table:bw_eff} the traffic allocation efficiency. We define this efficiency as the ratio between the required bandwidth of each flow, and the actual throughput, for each of the policies. Once more, we can see that congestion-aware policies are able to outperform the baseline operation, as all of them can effectively allocate more than the 90\% of the required bandwidth per flow for values between 2~Mbps and 6~Mbps. Also, notice how the performance of the two congestion-aware policies (i.e., SLCI and MCAA) is highly similar. 

\begin{figure}[t]
    \centering
    \includegraphics[width=\linewidth]{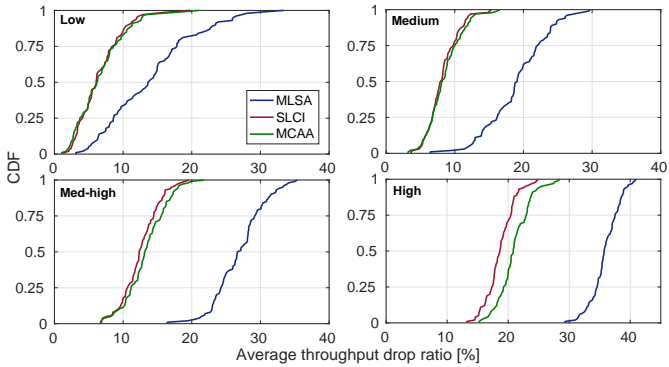}
    \caption{Drop ratio CDF for different network densities}\label{fig:dratio_cdf}
\end{figure}

Finally, we assess the performance of the different policies under high AP density deployments. To do so, we gradually increase the number of APs, to generate four different density deployments: low ($N=5$), medium ($N=10$), med-high ($N=20$) and high ($N=40$). For each AP, we placed a random number of stations (i.e., $M\sim U[15,25]$). Figure~\ref{fig:dratio_cdf} shows the empirical cumulative distribution function (CDF) of the average throughput drop ratio experienced for each density. The throughput drop ratio represents the percentage of traffic that cannot be served. Then, it is computed as one minus the ratio between the achieved throughput and the required one. As expected, the probability of having higher drop ratio values increases upon network densification. However, the different congestion-aware policies are able to reduce significantly those values. For instance, for medium density deployments, we find that drop ratio is decreased 2.25x by means of SLCI implementation for the 75th percentile compared to MLSA. Additionally, we find that SLCI further improves MCAA as the density increases. This phenomena occurs as MCAA policy allocates traffic to more interfaces, which in fact increases the exposure to the dynamics of the other APs. That is, in high density deployments, APs tend to have more neighboring APs, and so, the use of a high number of interfaces to transmit makes the traffic much more vulnerable as the probability of having a channel overlap increases.



\section{Conclusions \& Future Work}\label{Conclusions}

In this paper, we assessed and evaluated the adoption of multi-link communications by the upcoming IEEE 802.11be amendment. Through a wide variety of scenarios, our results show that the implementation of MLO can help the next generation Wi-Fi networks to satisfy the highly demanding requirements of modern applications. However, we show that the performance of MLO depends mainly on the implemented traffic allocation policy. From that side, as expected, congestion-aware allocation policies able to adapt to the instantaneous state of the network are the best performing ones. Additionally, we have observed that allocating an incoming flow to the emptiest interface is almost as good, if not better, than proportionally distributing the flow over multiple interfaces. Such conclusion relies on the fact that a single interface policy not only reduces traffic exposure, but minimizes the complexity of the traffic allocation procedure.

Regarding future work, we consider three potential extensions to further improve MLO operation. First, extend MCAA to re-allocate the traffic over the different interfaces periodically. Indeed, such implementation may minimize the negative impact of the activity of neighboring networks in terms of performance, but at the cost of entailing a much complex policy structure. Second, throughout this paper, we only considered IEEE 802.11be MLO capable networks, whereas in real scenarios they may coexist with legacy single link networks. Therefore, it may be relevant to study the different MLO traffic allocation policies under those conditions. Third, and last, to integrate the MLO operation in a Software Defined Networking framework \cite{lopez2019combining} to orchestrate the different APs that belong to the same WLAN, as well as to design and evaluate centralized multi-AP MLO policies.


\section*{Acknowledgments}

This  work  has  been  partially  supported by the Spanish Government under grant WINDMAL PGC2018-099959-B-I00 (MCIU/AEI/FEDER,UE), by the Catalan Government under grant 2017-SGR-1188, and  by a Gift from Cisco (2020).


\bibliographystyle{ieeetr}
\bibliography{main}
\end{document}